\documentclass{article}

\usepackage[a4paper, total={6.5in, 9in} ]{geometry}

\usepackage{amsmath} 
 
\usepackage{amsmath,amsfonts,amssymb}
\usepackage{graphicx}
\graphicspath{{figures/}}
\usepackage[colorlinks=true, allcolors=blue]{hyperref}
\usepackage{caption} 
\usepackage{booktabs}
\usepackage{multirow}
\usepackage{siunitx}

\usepackage{ctable}
\usepackage{makecell}
\usepackage{subfigure}
\usepackage{caption}
\usepackage{xcolor}
\usepackage{soul}
\usepackage{hyperref}
\hypersetup{
    colorlinks=true,
    linkcolor=blue,
    filecolor=magenta,      
    urlcolor=cyan,
}

\title{\LARGE \bf
Three-dimensional Segmentation of the Scoliotic Spine from MRI using Unsupervised Volume-based MR-CT Synthesis
}

\author{Enamundram M. V. Naga Karthik$^{1}$, Catherine Laporte$^{1,3}$, and Farida Cheriet$^{2,3}$
\renewcommand\footnotemark{}
\thanks{Accepted for Oral Presentation at the SPIE Medical Imaging Conference 2021, San Diego, CA.}
\thanks{$^{1}$Department of Electrical Engineering, \'Ecole de Technologie Sup\'erieure, Montr\'eal, Canada
        {\tt\small emvnagakarthik@gmail.com}}%
\thanks{$^{2}$Department of Computer Engineering, Polytechnique Montr\'eal, Montr\'eal, Canada}%
\thanks{$^{3}$CHU Sainte-Justine Research Center, Montr\'eal, Canada}%
}

\begin{document}

\maketitle
\thispagestyle{empty}
\pagestyle{empty}

\begin{abstract}

Vertebral bone segmentation from magnetic resonance (MR) images is a challenging task. Due to the inherent nature of the modality to emphasize soft tissues of the body, common thresholding algorithms are ineffective in detecting bones in MR images. On the other hand, it is relatively easier to segment bones from CT images because of the high contrast between bones and the surrounding regions. For this reason, we perform a cross-modality synthesis between MR and CT domains for simple thresholding-based segmentation of the vertebral bones. However, this implicitly assumes the availability of paired MR-CT data, which is rare, especially in the case of scoliotic patients. In this paper, we present a completely unsupervised, fully three-dimensional (3D) cross-modality synthesis method for segmenting scoliotic spines. A 3D CycleGAN model is trained for an unpaired volume-to-volume translation across MR and CT domains. Then, the Otsu thresholding algorithm is applied to the synthesized CT volumes for easy segmentation of the vertebral bones. The resulting segmentation is used to reconstruct a 3D model of the spine. We validate our method on 28 scoliotic vertebrae in 3 patients by computing the point-to-surface mean distance between the landmark points for each vertebra obtained from pre-operative X-rays and the surface of the segmented vertebra. Our study results in a mean error of 3.41 $\pm$ 1.06 mm. Based on qualitative and quantitative results, we conclude that our method is able to obtain a good segmentation and 3D reconstruction of scoliotic spines, all after training from unpaired data in an unsupervised manner. 

\end{abstract}

\section{Introduction}

Scoliosis is a complex three-dimensional (3D) deformity of the trunk. Along with a lateral deviation in the spine and the axial rotation of the vertebrae, a deformation of the rib cage is also observed. Severely afflicted patients are required to undergo surgical treatment to correct the spinal curvature. 
Intra-operative image guidance has also been suggested to improve surgical accuracy, but the high risk of radiation exposure precludes its common usage \cite{Chan2019}. In such cases, magnetic resonance imaging (MRI) acts as a reliable and radiation free pre-operative imaging evaluation, providing a 3D model of the spine, to which intra-operative images can be registered. Therefore, an accurate three-dimensional segmentation for creating 3D models of the vertebral column is an important step towards volume-guided surgical treatments. 
However, segmenting bones from MR images is a laborious task due to its poor contrast for bone structures. On the other hand, bones can be easily segmented in CT images, which provide much higher bone detail, albeit at the cost of exposing patients to harmful X-rays.
Therefore, we propose a method based on translating MR volumes to their corresponding CT volumes in an unsupervised fashion so that bone structures depicted in the MRI become easy to segment using a simple thresholding algorithm. 
A 3D CycleGAN model is trained for a completely unsupervised volume-to-volume translation across MR and CT domains. Based on the resulting synthesized CT volumes, the Otsu thresholding algorithm is used to segment the vertebral bones and finally obtain a 3D model of the vertebral column. The primary advantages of our method are that it is fully 3D, unsupervised and works with unpaired data, meaning that we do not need the corresponding CT volumes of the same patient for successful volume translation. This has important implications, especially in the case of scoliotic patients, where the risk of radiation of exposure is relatively high. 


\section{Related Work}
\label{sec:literature}

Vertebral bone segmentation in MR images is a well-studied problem, with approaches that can be classified into semi-automatic \cite{zukic-2012, purang-2014}, graph-based  \cite{egger_square-cut_2012,  schwarzenberg_cube-cut_2014} and learning-based \cite{chu_fully_2015, szu-hao_huang_learning-based_2009, Guerroumi2019}. However, there is a general lack of learning-based methods for the specific problem of segmenting scoliotic spines. 
This is primarily due to the unavailability of large and labeled spine datasets, let alone scoliotic spine datasets, for training.

Semi-automatic approaches rely on user-interaction for initialization of seed points on the vertebrae of interest.
Based on a single point-in-vertebra initialization, Zukic et al. \cite{zukic-2012} proposed a segmentation method in MR images using multiple feature boundary classification and iterative mesh inflation.  
Suzani et al. \cite{purang-2014} used an expectation maximization algorithm to align a statistical model from manually segmented CT volumes and user-initialized Canny edge-detected vertebrae. An important note it that seed initializations are done on a per-slice basis. 


Chu et al. \cite{chu_fully_2015} addressed the problem of localization and segmentation of vertebral bodies in 3D by using a unified random forest regression and classification framework.
Huang et al. \cite{szu-hao_huang_learning-based_2009} proposed an automatic method consisting of three stages: a modified AdaBoost algorithm for vertebral candidate detection followed by a refinement step via robust curve fitting, and a vertebrae segmentation using an iterative normalized-cut algorithm. In order to train the vertebrae detector, they manually labeled a set of 1398 images. 
Guerroumi et al. \cite{Guerroumi2019} and Neubert et al. \cite{Neubert2012} proposed automatic methods for simultaneously segmenting the vertebral bodies and interverterbral discs. While the latter performed a 3D segmentation using statistical shape analysis models, the former proposed a segmentation method from 2D MR images by incorporating a squeeze-and-excitation block in a U-Net-based network architecture.  Theirs is the first approach that performs segmentation on scoliotic spine images.
For our problem, manual segmentation is a major hindrance because each volume has about 216 2D slices on average and manually labeling each of them is impractical.  This provides strong motivation for an approach based on unsupervised learning.

Generative Adversarial Networks (GANs) \cite{goodfellow_generative_2014} have been widely used in many generative modelling tasks.
While the GAN-based literature \cite{Hiasa2018Grad, Armanious2019CycleMedgan, Zhang_2018_CVPR} mostly focuses on synthesizing other parts of the body across different imaging modalities, they nevertheless present an important step towards unsupervised cross-modality synthesis. To the best of our knowledge, there has not been any study yet that uses a GAN-based volume-translation approach for facilitating the three-dimensional segmentation of scoliotic spines. 

In image-to-image translation, images are mapped across two domains based on a pixel-to-pixel mapping. 
Conditional GANs \cite{isola_image--image_2018} do this in a setting where the images in the input and the output domains are paired.  
However, in the domain of medical images, paired datasets are rare. 
CycleGAN \cite{zhu_unpaired_2017} tackles this problem by translating images across domains with unpaired datasets. 
Modified versions of CycleGAN have recently been proposed to better suit the features of medical images. 
Specifically,  Hiasa et al.~\cite{Hiasa2018Grad} used gradient-consistency loss to improve the simulated image quality at the boundaries for their dataset containing MR and CT images of the pelvic region.
Armanious et al.~\cite{Armanious2019CycleMedgan} proposed the use of perceptual and style losses 
for PET-CT translation and MR motion correction.
Zhang et al. \cite{Zhang_2018_CVPR} employed a shape-consistency loss to constrain the geometric invariance of the synthetic data and performed volume-to-volume translation for segmenting 3D cardiovascular MR and CT data. Theirs is the first approach that uses a fully 3D method for segmenting multi-modal volumes.   

Considering our primary goal of 3D segmentation and volume reconstruction of the vertebral column, there are some major drawbacks in the literature: 
(1) methods using 2D images do not capture the spatial correlation between slices in volumetric data, 
(2) human supervision, be it in terms of manual segmentation of the ground truth labels or via user-interaction, is laborious and prevents 
scalability to a large number of volumes, and 
(3) a general lack of segmentation methods involving scoliotic spinal images. 
Therefore, we propose a method in this paper
that performs a volume-to-volume translation for facilitating the segmentation of scoliotic spines and reconstructs their 3D models, all without the need for labeled training data. 
The advantages of our method lie in the fact that the training is competely unsupervised and works with unpaired data.

\section{Methods}
\label{sec:methods}
This section describes our method for three-dimensional segmentation of scoliotic spines. We train a 3D CycleGAN model with gradient consistency loss from scratch to perform 
unsupervised volume-to-volume translation. The Otsu thresholding algorithm is then applied to segment bones from the synthesized CT volumes. Finally, a complete 3D model of the scoliotic spine is reconstructed based on the resulting segmentation. 

\begin{figure*}[htbp!]
    \centering
		\includegraphics[height=2.00in]{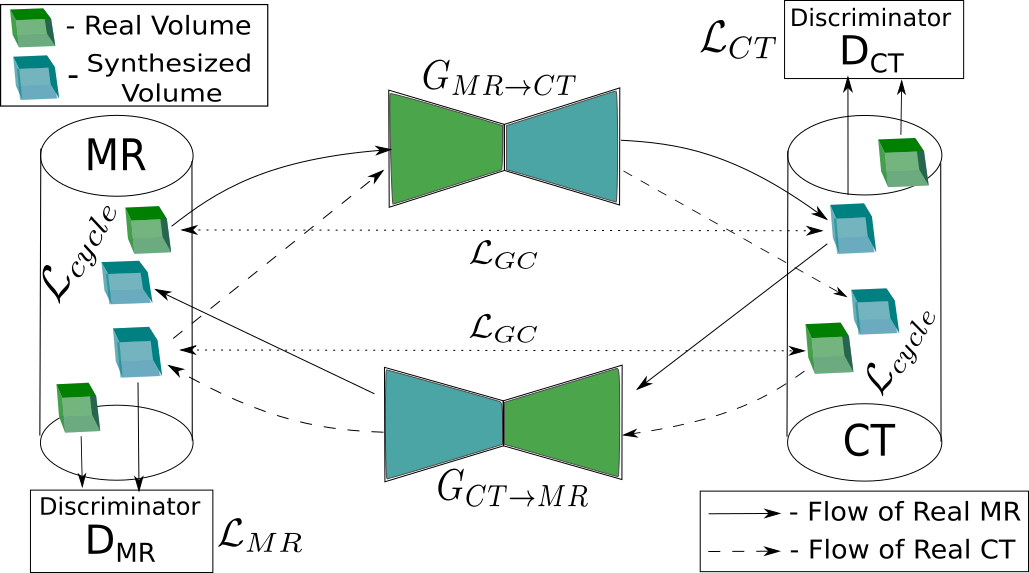}
	\caption[]{Each cylinder represents an imaging modality with green boxes showing the real volumes and blue boxes showing the synthetic volumes of that domain. \textit{Forward Cycle}: (shown by solid arrows) Starting from green box (left, top), going through $G_{MR \rightarrow CT}$ to blue box (right, top), then going through $G_{CT \rightarrow MR}$ to recover blue box (left, top).   \textit{Backward Cycle}: (shown by dashed arrows) Starting from green box (right, bottom), going through $G_{CT \rightarrow MR}$ to blue box (left, bottom), then going through $G_{MR \rightarrow CT}$ to recover blue box (right, bottom). The cycle-consistency loss ($\mathcal{L}_{Cycle}$) is calculated between the real and recovered volumes (top left and bottom right). The gradient-consistency loss ($\mathcal{L_{GC}}$) is calculated between the real and synthesized volumes (top left, top right and bottom left, bottom right). Figure adapted from \cite{Hiasa2018Grad}. }
	\label{fig-1}
\end{figure*}

Volume-to-volume translation is essentially a 3D extension of image-to-image translation to volumetric data. 
For a good segmentation, we found that it is best for the model to learn the correlation between the slices by itself. Hence, by using the 3D CycleGAN model, we propose to work directly on the volumes. The main idea is as follows: given a set of unpaired volumes from MR and CT domains, the model learns two function mappings simultaneously using two generators $G_{MR \rightarrow CT}$ and $G_{CT \rightarrow MR}$. Since voxel-wise comparison is infeasible due to the unavailability of paired data, the cycle-consistency loss is introduced as a comparison metric. 
Its idea: an input MR volume translated to the CT domain should ideally be recovered when that MR volume is translated back to the MR domain and vice-versa, that is, $ I_{MR} \approx G_{CT \rightarrow MR} (G_{MR \rightarrow CT} (I_{MR})) $ and $ I_{CT} \approx G_{MR \rightarrow CT} (G_{CT \rightarrow MR} (I_{CT})) $. In order for this reformulation to give reliable learning information, two discriminators $D_{MR}$ and $D_{CT}$ are used, whose task is to distinguish between the real volumes ($I_{MR}, I_{CT}$) and simulated 
volumes ($G_{CT \rightarrow MR}(I_{CT})$, $G_{MR \rightarrow CT}(I_{MR})$) respectively. Figure \ref{fig-1} illustrates the main idea.

\subsection{Materials}
\label{subsec:materials}

\begin{figure*}[htbp!]
    \centering
		\includegraphics[height=2.05in]{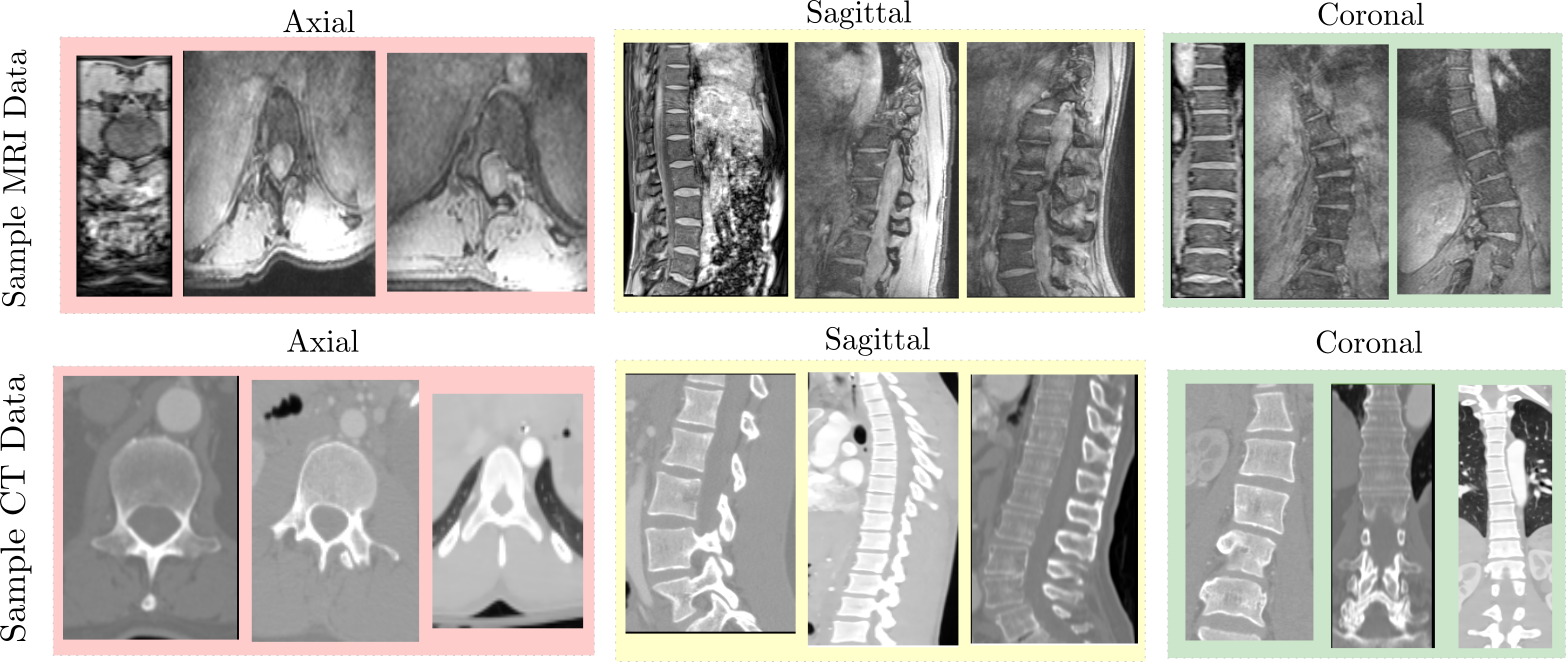}
	\caption[]{Sample axial, sagittal and coronal slices from 3 different input volumes each for MR and CT data. Note the contrast difference and the spinal curvature, highlighting the \textit{unpaired-ness} and a general lack of correspondence between the input MR and CT volumes.}
	\label{fig-2}
\end{figure*}

The MR and CT datasets were acquired from 2 different sources each. For MR, we used the dataset from the MICCAI 2018 Challenge on Automatic Intervertebral Disc Localization and Segmentation from 3D Multi-modality MR (M3) Images \footnote{https://ivdm3seg.weebly.com/}. This dataset consists of 16 volumes of the lumbar spine, comprising of 4 mutually aligned MR modalities. 
We chose to use only the water-phase images judging by a visual inspection of the contrast between the vertebral bodies and the surrounding regions. Our second source is a subset of the dataset described by Chevrefils et al. \cite{Chevrefils2009} consisting of MRI 3-D multi-echo data volumes from 11 adolescent idiopathic scoliotic (AIS) patients with deformities ranging from mild to severe, acquired from CHU Sainte-Justine in Montr\'eal, Qu\'ebec. This dataset focused only on the thoracic region 
of the vertebral column. 
For CT, we used 2 sample volumes provided by Slicer3D \footnote{https://www.slicer.org/}. These focused on the cardiac and the abdomen regions of the body, hence they were cropped accordingly to focus on the vertebral columns. 
To ensure uniformity across the data, all MR and CT volumes were cropped and resized to $256 \times 128 \times 48$ sized volumes. The voxel sizes were resized to lie in $\{1, 1.5\} \times \{1 \} \times \{1 \}$ $mm^3$ for MRI data and $\{0.75, 1\} \times \{0.75, 1\} \times \{1, 1.5\}$ $mm^3$ for CT data. 
Out of 27 (16 MICCAI + 11 Scoliotic) original MR volumes,  we had ground truth data for 3 scoliotic volumes obtained in the form of digitized landmarks reconstructed from preoperative biplanar X-ray scans. These 3 volumes were therefore used for testing. The rest of the volumes, along with their augmented versions, were used for training. The data augmentation was done offline, which included 3D rotation, Gaussian noise injection, elastic deformation, and contrast stretching methods. The Gaussian noise was sampled from a normal distribution with mean ($\mu$) $0$ and standard deviation ($\sigma$) $0.01$. In total, 1104 MR volumes and 200 CT volumes were used for training. Figure \ref{fig-2} shows a few instances of the training data.

\subsection{Unsupervised Volume-to-Volume Translation}
\label{subsec:translation}
For accurate thresholding-based segmentation of the bones from synthesized CT volumes, the emphasis should be placed on translating the boundaries of the vertebral bodies correctly, as opposed to their insides.
Therefore, we use gradient correlation (GC), a commonly used image-similarity metric in 2D-3D medical image registration. It is defined as the normalized cross correlation (NCC) between the gradients of two images. Given the difference in the voxel intensities of bones in MR and CT volumes, maximizing cross correlation between them helps in filtering low frequency differences  (such as soft tissues) and the inherent nature of image-gradients acting as a high pass filter, focuses on registering the bone anatomy that is prominent in CT volumes. GC similarity has shown promising results in the context of vertebral level localization\cite{De_Silva_2016}. 

Our model consists of three types of loss terms: the adversarial losses ($\mathcal{L}_{CT}$ and $\mathcal{L}_{MR}$), the cycle-consistency loss ($\mathcal{L}_{Cycle}$) and the gradient consistency loss ($\mathcal{L}_{GC}$). 
\begin{equation}
G_{CT \to MR}^*, G_{MR \to CT}^* = \arg \min_{ \underset{  G_{CT \to MR} } { G_{MR \to CT}  }  } \max_{  \underset{ D_{CT} }{ D_{MR}}  }  ( \mathcal{L}_{CT} + \mathcal{L}_{MR} + \lambda \mathcal{L}_{Cycle} + \gamma \mathcal{L}_{GC} ),
\end{equation}
where $\lambda$ and $\gamma$ are the hyperparameters for weighting cycle- and gradient-consistency losses. We set $\lambda=10.0$ and $\gamma = 0.1\lambda$ for this study.

The adversarial loss helps in mapping the source data distribution to the target data distribution. For the mapping defined by $ G_{MR \rightarrow CT}: I_{MR} \to I_{CT} $ and its discriminator $ D_{CT} $, the objective is defined as
\begin{equation}
	\mathcal{L}_{CT} = \mathbb{E}_{x \sim I_{CT}} [ \log D_{CT}(x) ]   +  \mathbb{E}_{y \sim I_{MR}} [ \log ( 1 - D_{CT}(G_{MR \to CT}(y) ) ) ].
\end{equation}
Similarly, the objective for the reverse path is defined as:
\begin{equation}
	\mathcal{L}_{MR} = \mathbb{E}_{y \sim I_{MR}} [ \log D_{MR}(y) ]   +  \mathbb{E}_{x \sim I_{CT}} [ \log ( 1 - D_{MR}(G_{CT \to MR}(x) ) ) ],
\end{equation}
where $x$ and $y$ are the volumes from CT and MR domains respectively. The reader is referred to figure \ref{fig-1} for visual illustration.

The generators are enforced to be cycle-consistent in order to learn the function mappings with \textit{unpaired} data. 
To steer the learning towards this constraint, the cycle-consistency loss is defined as:
\begin{equation}
	\begin{split}
		\mathcal{L}_{Cycle}  & =  \mathbb{E}_{x \sim I_{CT}} [ || G_{MR \rightarrow CT} (G_{CT \rightarrow MR} (x) )  -  x ||_1  ]  +  \mathbb{E}_{y \sim I_{MR}} [ || G_{CT \rightarrow MR} (G_{MR \rightarrow CT} (y) )  -  y ||_1  ],
	\end{split}
\end{equation}
where $ || \bullet ||_1$ denotes the $L_1$ norm between the real and recovered volumes for each domain.

As mentioned earlier, we used GC to enforce edge similarity between real and synthesized volumes and filter out low-frequency differences. Therefore, GC between two volumes is defined as:
\begin{equation}
GC(A, B) = \frac{1}{2} \left( NCC (\nabla_x A, \nabla_x B) + NCC (\nabla_y A, \nabla_y B) + NCC (\nabla_z A, \nabla_z B)  \right),
\end{equation}
where $ 
NCC (\nabla_x A, \nabla_x B) = \left( \frac{ \sum_{i,j} (\nabla_x A - \mu_{\nabla_x A}) (\nabla_x B - \mu_{\nabla_x B})}{ \sqrt{\sum_{i,j} (\nabla_x A - \mu_{\nabla_x A})^2 }  \sqrt{\sum_{i,j} (\nabla_x A - \mu_{\nabla_x A})^2 }  } \right)
$ 
and $\nabla_x, \nabla_y, \nabla_z$ are the gradients of the input volume in $x, y$ and $z$ axes respectively. $\mu_{\nabla_i J}$ is the mean of the gradient of volume $J$ in axis $i$. Therefore, gradient consistency loss is defined as:
\begin{equation}
	\mathcal{L}_{GC}  = \frac{1}{2} \left[ \mathbb{E}_{x \sim I_{CT}} (1 - GC(x, G_{CT \rightarrow MR} (x)))  +  \mathbb{E}_{y \sim I_{MR}} (1 - GC(y, G_{MR \rightarrow CT} (y)))  \right].
\end{equation}

\subsection{Otsu Thresholding and Volume Reconstruction}
\label{subsec:threshold}
Otsu's method is a widely used image thresholding algorithm that maximizes inter-class pixel intensity variance and returns a single intensity threshold, thus  separating the image into foreground and background pixels. We used the Otsu thresholding option provided in the ``Segment Editor'' module by Slicer3D. Based on a tunable range of pixel intensity values, the thresholding algorithm is applied on a per-slice basis. Finally, we used the median smoothing operation which reduces noise while preserving the edges. The resulting post-processed segmentation was used for generating a 3D model of the scoliotic spine. The outputs at each stage of the method are shown in figure \ref{fig-3}. 


\section{Results}
\label{sec:results}

\begin{figure*}[t]
    \centering
		\includegraphics[height=3.25in]{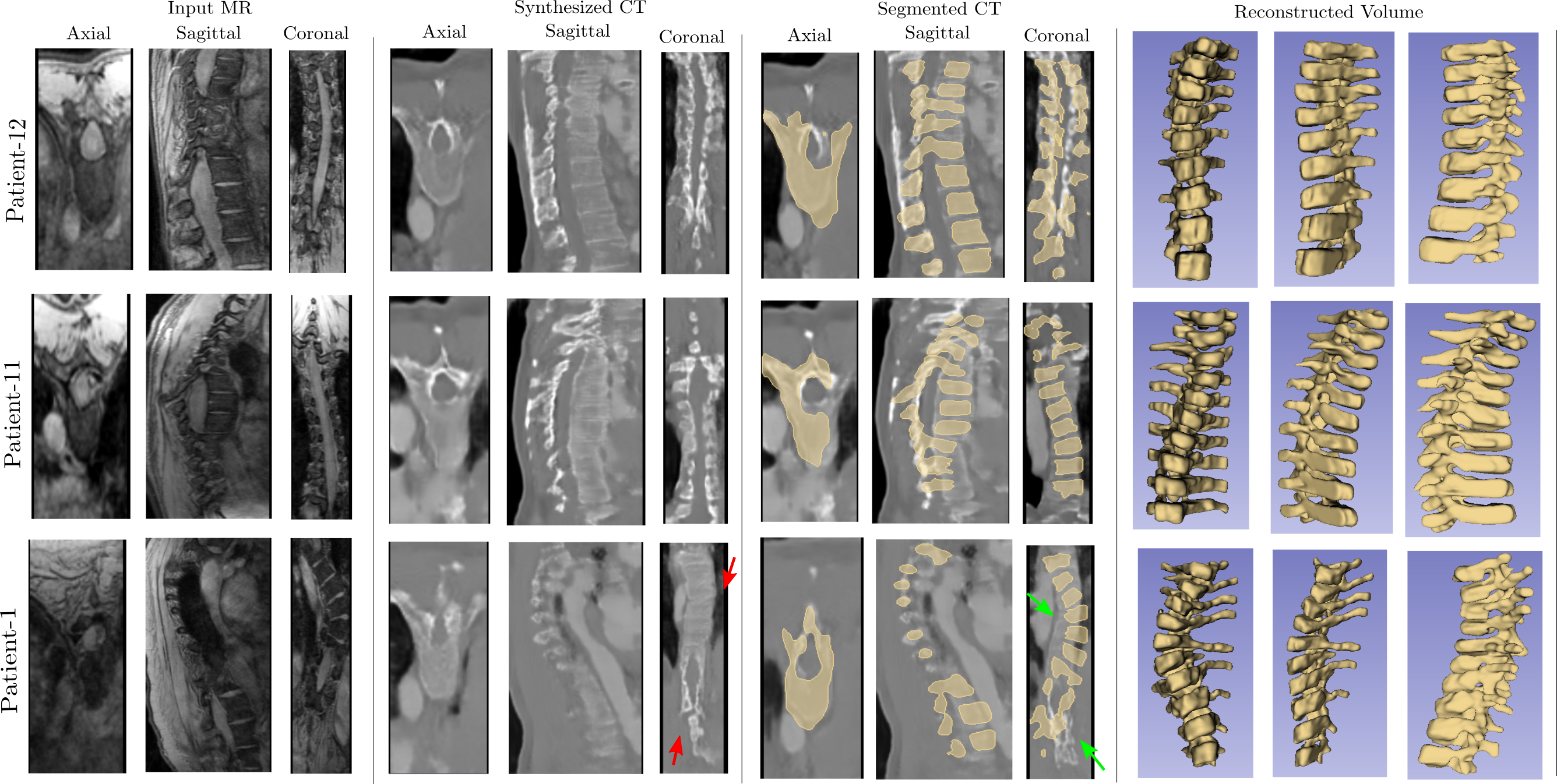}
	\caption[]{3D segementation and volume reconstruction of the vertebral column. Top $\to$ Bottom: Results corresponding to the scoliotic spines of Patients 12, 11, and 1 respectively. Right $\to$ Left: First column shows 2D slices of MR volumes, Second column shows the synthesized CT volumes, Third column shows the result of 3D segmentation by Otsu thresholding + post-processing, Fourth column shows the reconstructed 3D models of scoliotic spines. The red arrows show the translation errors and the green arrows show the corrected segmentations.}
	\label{fig-3}
\end{figure*}

In this section, we present both qualitative and quantitative analysis of our model's performance on a test data of 3 scoliotic patients (P1, P11 and P12). Figure \ref{fig-3} shows the qualitative results obtained at each stage. For the regions of the spine where the curvature is normal, we found that the model learned to identify the difference in the intensities between bones and the soft tissues in MR and CT volumes and is able to translate them (figure \ref{fig-3}, Patient-12 and Patient-11). 
However, some incorrect translations in the synthesized CT volumes were also observed. This was prominent in P1's case (figure \ref{fig-3}, third row, red arrows) where the spinal curvature was too high for a good translation. As a result, the subsequent segmentation errors due to Otsu thresholding were manually corrected using Slicer's paint-brush tool (figure \ref{fig-3}, third row, green arrows). 
This is attributed to the unpaired nature of the input data and a relatively high anatomical variation in the scoliotic spine. Even though the segmentation is not completely automatic due to translation errors, post-processing of the segmentation resulting from synthesized CT volumes is certainly easier compared to a direct segmentation from MR volumes. 
Furthermore, we note that even in the case of real CT volumes, some manual intervention 
was necessary in order to remove the segmented structures other than the spine after thresholding.  


\begin{figure*}[htbp!]
    \centering
		\includegraphics[width=6.25in]{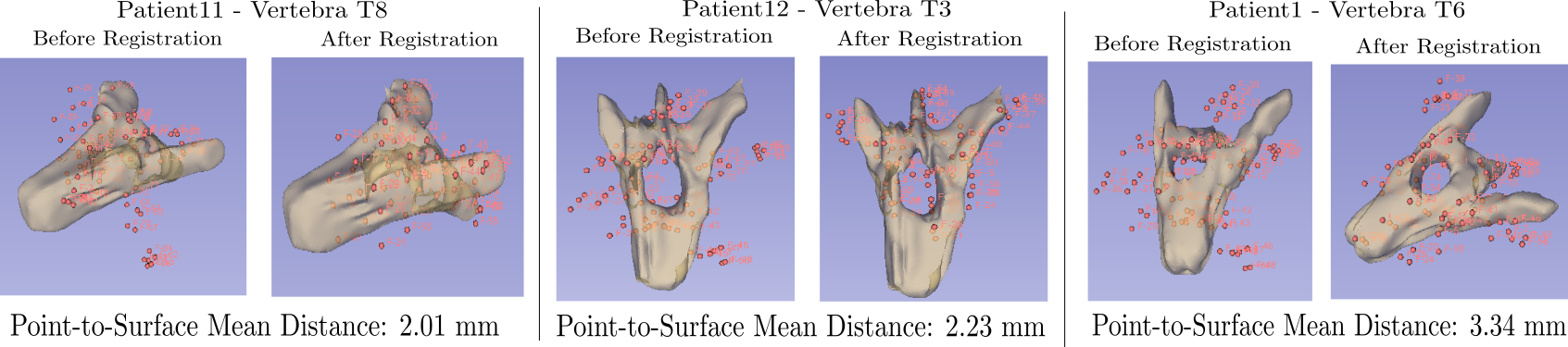}
	\caption[]{Anatomical landmarks (in pink) overlayed on individually segmented vertebra. Notice that before registration, mainly the spinous and transverse processes are not aligned. A rigid transformation, followed by ICP registration, aligns the two shapes and the mean distance between them is computed.}
	\label{fig-4}
\end{figure*}

To evaluate the accuracy of the 3D reconstructed vertebrae (figure \ref{fig-3}, last column), we compared the anatomical landmarks from the corresponding biplanar X-ray data sets and the segmented vertebrae by calculating the point-to-surface mean distance \cite{Delorme2003}. The thoracic vertebrae under consideration, T2-T11, had 74 labeled landmark points. The preoperative biplanar X-rays were taken with the patient in standing position whereas the MRIs were acquired with the patient in supine position, causing significant differences between the overall spine shapes. However, the individual vertebrae within the spine remain rigid irrespective of the patient's posture. Hence, instead of validating on the entire vertebral column, we calculated the point-to-surface distance for each vertebra. The iterative closest point (ICP) algorithm was used for registration of corresponding vertebrae before computing the mean distance error. Figure \ref{fig-4} shows an example of the anatomical landmarks overlayed on the segmented vertebra before and after performing the registration. 
In order to ensure that ICP has a good initialization, the landmark points were roughly aligned \textit{a priori} by a rigid transformation. The ICP algorithm was then run for 100 iterations and the point-to-surface mean distance was computed. Table \ref{table-1} shows the resulting point-to-surface errors for different vertebral levels.
Out of the 3 test patients, best results are obtained from the reconstructed vertebrae of patients 11 and 12 resulting in an average error of less than 3 mm despite the error of the T2 vertebra being high. The poor results for the T2 vertebra are explained by our model's inability to translate image intensities present at the top and bottom regions. Since a part of our MR dataset focuses on the T2-T11 region, training the model on a longer set of vertebrae with a few possible buffer vertebrae at the top and bottom, could improve the translation of the specific vertebrae under consideration.  
We also observed systematically high errors at all vertebral levels in patient 1. This is due to a relatively higher spinal curvature, especially in the T2-T6 region (figure \ref{fig-3}, third row, first column) caused by a more severe deformity.  Such high anatomical variation resulted in a poor synthesis of the corresponding CT volume, hence affecting the subsequent Otsu thresholding and 3D reconstruction. 
Our results are comparable (with less variance, however) to Delorme et al. \cite{Delorme2003} who reported a global 3D reconstruction error of 3.3 $\pm$ 3.8 mm for 60 scoliotic vertebrae using the  landmarks 2D radiographic images as ground truth. 
While a similar magnitude of error is highly encouraging, it also suggests that at least some of our error inherently accounts for a possible error in the digitization of these ground truth landmarks, and therefore is not solely based on our method of vertebrae reconstruction. 

\begin{table}
\centering
\caption{Accuracy of the 3D reconstructions of the vertebrae shown for $28$ Scoliotic vertebrae in 3 patients.}
  \begin{tabular}{cSSSS}
    \toprule
    \multirow{1}{*}{Level} &
      \multicolumn{3}{c}{Point-to-surface mean distance (mm)} &
      \multicolumn{1}{c}{Mean per Level (mm)} \\
      & {Patient 12}  & {Patient 11} &  {Patient 1} \\
      \midrule
    T2 & 3.68 & 5.42 & 5.78  & 4.96 \\
    T3 & 2.23 & 3.23 & 5.48  & 3.65 \\
    T4 & 2.56 & 2.84 & 3.67  & 3.02 \\
    T5 & 3.02 & 3.12 & 3.61  & 3.25 \\
    T6 & 2.73 & 2.76 & 3.34  & 2.94 \\
    T7 & 2.77 & 2.23 & 5.47  & 3.49 \\
    T8 & 2.64 & 2.01 & 4.59  & 3.08 \\
    T9 & 3.07 & 2.82 & 3.31  & 3.07 \\
    T10 & 3.15 & 2.34 & 4.81 & 3.43 \\
    T11 & $N/A$ & 2.89 & $N/A$  & 2.89 \\
    \midrule
    Mean $\pm$ S.D & \hspace{11mm} \text{2.86 $\pm$ 0.39} & \hspace{12mm} \text{2.97 $\pm$ 0.90} & \hspace{11mm} \text{4.45 $\pm$ 0.93}  \\
    \midrule
    \multicolumn{1}{c}{\textbf{Total}} & 
    \multicolumn{4}{c}{\textbf{28 Vertebrae} with a mean distance error of \textbf{3.41 $\pm$ 1.06 mm}} \\
    \bottomrule
  \end{tabular}
  \label{table-1}
\end{table}

\section{New Work to be Presented}
We present an unsupervised and fully 3D method for segmenting the vertebral column in scoliotic patients. 
Our method makes the laborious task of bone segmentation from MR volumes easy by synthesizing their corresponding CT volumes from an unpaired volume-to-volume translation. 
Acquiring paired MR and CT datasets of scoliotic patients is extremely difficult, especially considering the high risk of radiation exposure. In such cases, our method helps in reconstructing a complete 3D model of the scoliotic spine given that patient's MRI volume. 
A 3D model of the spine can serve as a useful tool in surgical planning and also improve the current clinical procedure for the follow-up of scoliosis, which requires patients to undergo several X-ray examinations for assessing the spinal curvature. 

\section{Conclusions}
\label{sec:conclusions}
A fully three-dimensional method for segmenting scoliotic spines was presented. The method performs unsupervised volume-to-volume translation for a 3D cross-modality synthesis using unpaired MR and CT volumes of the spine, allowing Otsu thresholding to be effectively applied to the 
synthesized CT volumes for segmentation and a 3D model to be reconstructed. 
The point-to-surface mean distance is computed between ground truth landmarks and 3D reconstructions for $28$ scoliotic vertebrae, resulting in a mean error of 3.41 $\pm$ 1.06 mm, which is on a similar order of magnitude to the accuracy of the ground truth measurements. 
Qualitative results show that our method is easily able to segment vertebral bones from synthesized CT volumes by training on unpaired data instances with no expert supervision. 
Future work would be directed towards making this method completely automatic, thus enabling it to scale to a large number of volumes without requiring any manual intervention. Another direction would be to validate our method on a larger test dataset and also on manually segmented MR volumes, which act as a higher quality gold standard.

\section*{Acknowledgements}
This work was funded by the Natural Science and Engineering Research Council of Canada.
We thank Compute Canada \footnote{https://www.computecanada.ca/home/} for providing the computational resources used in this work.

\bibliographystyle{unsrt}
\bibliography{references_final}

\end{document}